\documentclass[12pt,showpacs,prd]{revtex4}
\usepackage{epsfig,amssymb,amsfonts}

\makeatletter
\renewcommand{\@makefntext}[1]{\parindent=1em\noindent\hbox to 1.8em{\hss$^{\@thefnmark}$}#1}
\renewcommand{\@footnotemark}{\hbox{\mathsurround=0pt$^{\@thefnmark}$}}
\newcommand{\ftnote}[2]{\footnotemark[#1]\footnotetext[#1]{#2}}
\makeatother

\DeclareMathSymbol{\varGamma}{\mathord}{letters}{"00}

\begin{document}
\title{On Goldstone bosons decoupling from high--lying hadrons}
\author{L. Ya. Glozman}
\affiliation{Institute for 
Physics, University of Graz, Universit\"atsplatz 5, A-8010
Graz, Austria}

\author{A. V. Nefediev}
\affiliation{Institute of Theoretical and Experimental
Physics, 117218, B. Cheremushkinskaya 25, Moscow, Russia}

\newcommand{\be}{\begin{equation}}
\newcommand{\bea}{\begin{eqnarray}}
\newcommand{\ee}{\end{equation}}
\newcommand{\eea}{\end{eqnarray}}
\newcommand{\ds}{\displaystyle}
\newcommand{\low}[1]{\raisebox{-1mm}{$#1$}}
\newcommand{\loww}[1]{\raisebox{-1.5mm}{$#1$}}
\newcommand{\lmn}{\mathop{\sim}\limits_{n\gg 1}}
\newcommand{\vpint}{\int\makebox[0mm][r]{\bf --\hspace*{0.13cm}}}
\newcommand{\too}{\mathop{\to}\limits_{N_C\to\infty}}
\newcommand{\vp}{\varphi}
\newcommand{\vx}{{\vec x}}
\newcommand{\vy}{{\vec y}}
\newcommand{\vz}{{\vec z}}
\newcommand{\vk}{{\vec k}}
\newcommand{\vq}{{\vec q}}
\newcommand{\vpp}{{\vec p}}
\newcommand{\vn}{{\vec n}}
\newcommand{\vg}{{\vec \gamma}}

\begin{abstract}
In this paper, we discuss a decoupling of the Goldstone bosons
from highly excited hadrons in relation to the restoration
of chiral symmetry in such hadrons. We use a generalised
Nambu--Jona-Lasinio model with the interaction between quarks in
the form of an instantaneous Lorentz--vector confining potential.
This model is known to provide spontaneous breaking of chiral
symmetry in the vacuum via the standard selfenergy loops for
valence quarks. For highly excited hadrons, where the typical
momentum of valence quarks is large, the loop contributions
represent only a small correction to the chiral--invariant
classical contributions and asymptotically vanish.
Consequently the chiral symmetry violating Lorentz--scalar
dynamical mass of quarks vanishes. Then the conservation
of the axial vector current in the chiral limit requires, via the
Goldberger--Treiman relation, that the valence quarks decouple
from the Goldstone boson. As a consequence, the whole hadron decouples
from the Goldstone boson as well, which implies that its axial constant
also vanishes.
\end{abstract}
\pacs{12.38.Aw, 12.39.Ki, 12.39.Pn}
\maketitle

\section{Introduction}

An approximate restoration of $SU(2)_L \times SU(2)_R$ and $U(1)_A$ symmetries
in excited hadrons has recently become a subject of a significant
theoretical effort \cite{G1,CG1,G2,G3,G4,G5,OPE,SWANSON,DEGRAND,SHIFMAN,parity2,
GNR,JPS,CG2}. This {\em effective}  restoration of  chiral symmetry requires in
particular that highly excited hadrons should gradually
decouple from the Goldstone bosons \cite{G3,JPS,CG2}. There is
an indirect phenomenological hint for such a decoupling.
Indeed, the coupling constant for the process $h^*\to h+\pi$ decreases 
for high lying resonances, because the phase--space factor for such a decay
increases with the mass of a resonance much faster than the decay width. 

A coupling of the Goldstone bosons to the valence quarks is
regulated by the conservation of the axial current (we consider
for simplicity the chiral limit). This conservation results in
a Goldberger--Treiman relation \cite{GT}, taken at the ``constituent quark" level, giving
\be
g_\pi\propto m_q^{\rm eff},
\label{GT}
\ee
where $m_q^{\rm eff}$ is the quark Lorentz--scalar dynamical mass which
appears selfconsistently due to spontaneous breaking of chiral symmetry (SBCS) in the vacuum. 
Appearance of such a dynamical mass is a general feature of chiral symmetry 
breaking and has been studied in great detail in the context
of different models like the Nambu and Jona-Lasinio model
\cite{NJL,REVIEWS}, instanton liquid model \cite{SHURYAK}, within
the Schwinger--Dyson formalism with the quark kernel formed by the QCD string \cite{ST}, by the perturbative gluon
exchange \cite{Roberts} or by the instantaneous Lorentz--vector confinement
\cite{Orsay,Orsay2,Lisbon}. A general feature of this dynamical mass
is that it results from quantum fluctuations of the quark
field and vanishes at large momenta where the classical contributions
dominate \cite{G4,GNR}. Then, since the average momentum
of valence quarks higher in the spectrum increases, the valence
quarks decouple from the quark condensate and their dynamical
Lorentz--scalar mass decreases (and asymptotically vanishes), so that 
chiral symmetry is {\em approximately} restored in the highly--excited hadrons \cite{G1,parity2,GNR}. This
implies, via the Goldberger--Treiman relation (\ref{GT}),
that valence quarks, as well as the whole hadron, decouple 
from the Goldstone bosons \cite{G3}. This, in turn, requires 
the axial coupling constant of the highly excited hadrons to decrease
and to vanish asymptotically.

While this perspective was shortly outlined in the past, this
has never been considered in detail microscopically. However, it is important
to clarify this physics, especially because the origins of
this phenomenon cannot be seen at the level of the effective
Lagrangian approach, where the coupling constant of the Goldstone
boson to the excited hadron is an input parameter and the
decoupling looks unintuitive \cite{JPS}.

Even though the role of different gluonic interactions in QCD, which
could be responsible for chiral and $U(1)_A$ symmetries breaking, 
is not yet clear, the most fundamental reason for the restoration
of these symmetries in excited hadrons is universal \cite{G4}.
Namely, both $SU(2)_L \times SU(2)_R$ and $U(1)_A$ symmetries
breaking result from quantum fluctuations of the quark fields
(that is, loops). However, for highly-excited hadrons, where the
action of the intrinsic motion is large, a semiclassical regime
necessarily takes place. Semiclassically, the contribution of
quantum fluctuations is suppressed, relative to the classical
contributions, by a factor $\hbar/{\cal S}$, where ${\cal S}$ is the classical
action of the intrinsic motion in the  hadron in terms of the quark 
and gluon degrees of freedom. Since, for highly excited hadrons,
${\cal S} \gg \hbar$, contributions of the quantum fluctuations
of the quark fields are suppressed relative to the classical 
contributions. Consequently both chiral and $U(1)_A$ symmetries are
approximately restored in this part of the spectrum.

Although this argument is quite general and solid, it does not
provide one with any detailed microscopic picture of the
symmetry restoration. Then in the absence of controllable
analytic solutions of QCD such an insight can be obtained
only through models. 

It is instructive to outline the minimal set of requirements
for such a model. It must be (i) relativistic, (ii) chirally
symmetric, (iii) able to provide spontaneous breaking of chiral
symmetry, (iv) it should contain confinement, (v) it must explain
the restoration of chiral symmetry in excited states. There is a
model which does incorporate all required elements. This is the
generalised Nambu and Jona-Lasinio (GNJL) model with the instantaneous
Lorentz--vector confining kernel \cite{Orsay,Orsay2,Lisbon}. 
In this model, confinement of quarks is guaranteed due to instantaneous infinitely rising
(for example, linear) potential. Then chiral symmetry breaking can be described
by the standard summation of the valence--quark selfinteraction loops giving rise to the
Schwinger--Dyson equation for the quark selfenergy \cite{Orsay,Orsay2}.
Alternatively the model can be considered in the Hamiltonian approach using the BCS formalism \cite{BCS}. In this case,
chiral symmetry breaking in the vacuum happens via condensation of the ${}^3P_0$ quark--antiquark pairs and dressed quarks appear from the
Bogoliubov--Valatin transformation applied to bare quarks. The mass--gap equation ensures absence of anomalous Bogoliubov terms
in the Hamiltonian \cite{Lisbon}. Finally, mesons are built using the Bethe--Salpeter equation for the quark--antiquark bound states
\cite{Orsay2,Lisbon} or by a generalised bosonic Bogoliubov-like transformation applied to the operators creating/annihilating 
quark--antiquark pairs \cite{NR1}.

It was demonstrated in Ref.~\cite{parity2} that, for the low--lying states,
where the typical momentum of valence quarks is not large and chiral
symmetry breaking is important, this model leads to an {\it effective}
Lorentz--scalar binding potential, while for high--lying states, such an effective
potential becomes a pure Lorentz spatial vector. Then the discussed above quantum nature of chiral symmetry breaking in QCD
as well as the transition to the semiclassical regime for excited states, with loop effects being suppressed relative
to the classical contributions, has been illustrated within the same model in Ref.~\cite{GNR}.
As a result, the model does provide chiral symmetry restoration for excited hadrons. 

The purpose of this paper to give an insight into physics of decoupling of the Goldstone bosons from the
excited hadrons which happens in line, and due to the same reason, with the approximate restoration of chiral symmetry for these hadrons. 
We resort to the GNJL model in view of its obvious advantage as a tractable model for QCD which can be used as a laboratory to get
a microscopical insight into the restoration of chiral symmetry for excited hadrons.

\section{Generalised Nambu--Jona-Lasinio model}

\subsection{Some generalities}

In this chapter, we overview the GNJL chiral quark model \cite{Orsay,Orsay2,Lisbon} which is described by the Hamiltonian 
\be 
\hat{H}=\int d^3x\bar{\psi}(\vec{x},t)\left(-i\vec{\gamma}\cdot
\vec{\bigtriangledown}+m\right)\psi(\vec{x},t)+ \frac12\int d^3
xd^3y\;J^a_\mu(\vec{x},t)K^{ab}_{\mu\nu}(\vec{x}-\vec{y})J^b_\nu(\vec{y},t),
\label{H} 
\ee 
with the quark current--current ($J_{\mu}^a(\vec{x},t)=\bar{\psi}(\vec{x},t)\gamma_\mu\frac{\lambda^a}{2}
\psi(\vec{x},t)$) interaction parametrised by an instantaneous confining kernel $K^{ab}_{\mu\nu}(\vec{x}-\vec{y})$ of a generic
form. In this paper, we use the simplest form of the kernel compatible with the requirement of confinement,
\be
K^{ab}_{\mu\nu}(\vec{x}-\vec{y})=g_{\mu 0}g_{\nu 0}\delta^{ab}V_0(|\vec{x}-\vec{y}|).
\label{KK}
\ee
We do not dwell at any particular form of the confining potential $V_0(|\vec{x}-\vec{y}|)$, though,
if needed for an illustration purpose, we employ a power-like confining potential \cite{Orsay,replica4},
\be 
V_0(|\vec{x}|)=K_0^{\alpha+1}|\vec{x}|^{\alpha},\quad 0\leqslant\alpha\leqslant 2,
\label{potential} 
\ee
for qualitative analysis, concentrating mostly at the case of the linear confinement ($\alpha=1$) or,
for numerical studies, resorting to the harmonic oscillator potential ($\alpha=2$) \cite{Orsay,Orsay2,Lisbon}.

\subsection{Chiral symmetry breaking}

\begin{figure}[t]
\begin{center}
\epsfig{file=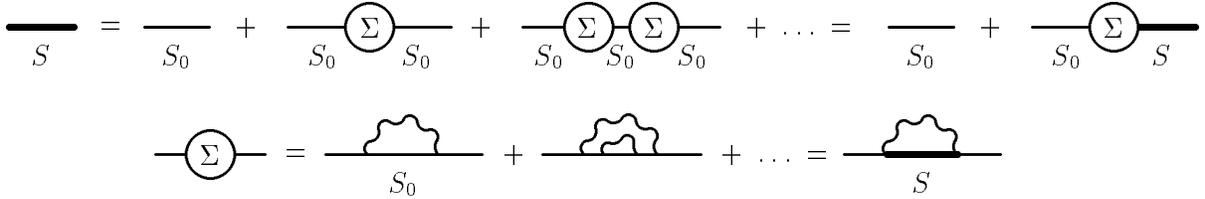,width=16cm} \caption{Graphical
representation of the equations for the dressed quark propagator,
Eq.~(\ref{Ds}), and for the quark mass operator,
Eq.~(\ref{Sigma01}).}\label{diagrams}
\end{center}
\end{figure}

Spontaneous breaking of chiral symmetry in the class of Hamiltonians (\ref{H}) is described via the standard
Dyson series for the quark propagator which takes the form, schematically (see Fig.~\ref{diagrams}): 
\be 
S=S_0+S_0\Sigma S_0+S_0\Sigma S_0\Sigma S_0+\ldots=S=S_0+S_0\Sigma S, 
\label{Ds} 
\ee 
with $S_0$ and $S$ being the bare-- and the dressed--quark propagators,
respectively, $\Sigma$ is the quark mass operator.
The Dyson--Schwinger Eq.~(\ref{Ds}) has the solution
\be
S^{-1}(p_0,\vpp)=S_0^{-1}(p_0,\vpp)-\Sigma(\vpp), 
\label{Sm1} 
\ee
where the mass operator independence of the energy $p_0$ follows from the instantaneous nature of the interaction. 
The expression for the mass operator through the dressed--quark propagator (see Fig.~\ref{diagrams}) reads: 
\be
i\Sigma(\vec{p})=C_F\int\frac{d^4k}{(2\pi)^4}V_0(\vec{p}-\vec{k})\gamma_0 S(k_0,\vec{k})\gamma_0,\quad 
C_F=\frac{N_C^2-1}{2N_C},
\label{Sigma01} 
\ee 
with both quark--quark--potential vertices being bare momentum--independent vertices $\gamma_0$. This
corresponds to the so-called rainbow approximation which is well justified in the limit of the large number of colours $N_C$. We
assume this limit in what follows. Then all nonplanar diagrams
appear suppressed by $N_C$ and can be consecutively removed from the theory. Eqs.~(\ref{Sm1}) and (\ref{Sigma01}) together produce 
a closed set of equations, equivalent to a single nonlinear equation for the mass operator, 
\be
i\Sigma(\vec{p})=\int\frac{d^4k}{(2\pi)^4}V(\vec{p}-\vec{k})
\gamma_0\frac{1}{S_0^{-1}(k_0,\vk)-\Sigma(\vk)}\gamma_0, 
\label{Sigma03} 
\ee 
where the fundamental Casimir operator $C_F$ is absorbed by the potential,
$V(\vpp)=C_FV_0(\vpp)$.

To proceed, we use the standard parametrisation of the mass operator $\Sigma(\vpp)$ in the form: 
\be
\Sigma(\vec{p})=[A_p-m]+(\vec{\gamma}\hat{\vec{p}})[B_p-p],
\label{SiAB} 
\ee 
so that the dressed--quark Green's function (\ref{Sm1}) becomes 
\be
S^{-1}(p_0,\vec{p})=\gamma_0p_0-(\vec{\gamma}\hat{\vec{p}})B_p-A_p,
\label{SAB} 
\ee 
where, due to the instantaneous nature of the interquark interaction, the time component of the four--vector
$p_\mu$ is not dressed.

It is easily seen from Eq.~(\ref{SAB}) that the functions $A_p$ and $B_p$ represent the scalar and the space--vectorial part
of the effective Dirac operator, respectively. Notice, that it is the scalar
part $A_p$ that breaks chiral symmetry and hence it can be
identified with the dynamical mass $m_q^{\rm eff}$ of the valence quark appearing in the Goldberger--Treiman relation (\ref{GT}).
In the chiral limit, $A_p$ vanishes, unless chiral symmetry is broken spontaneously. It is convenient, therefore, to introduce an angle, 
known as the chiral angle $\vp_p$, according to the definition: 
\be
\tan\vp_p=\frac{A_p}{B_p}, 
\label{cha} 
\ee 
and varying in the range $-\frac{\pi}{2}<\vp_p\leqslant\frac{\pi}{2}$, with the
boundary conditions $\vp(0)=\frac{\pi}{2}$, $\vp(p\to\infty)\to 0$.

\begin{figure}[t]
\begin{center}
\epsfig{file=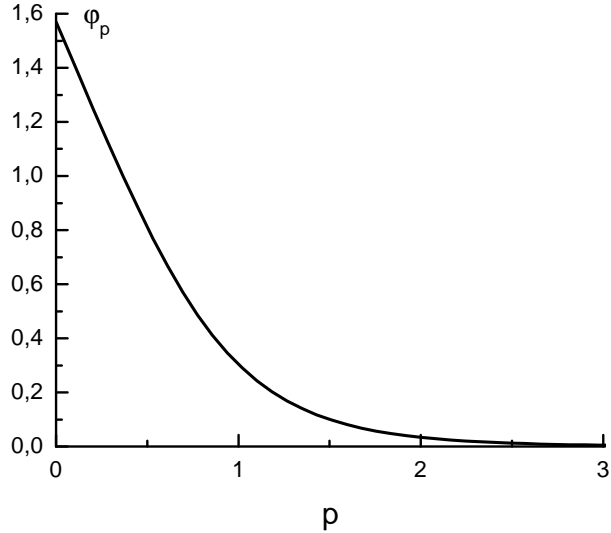,width=8cm} \caption{Nontrivial solution to the
mass--gap Eq.~(\ref{mge2}) with $m=0$ and for the linear
confinement (see, for example, Refs.~\cite{replica1,replica4} for the details). The
momentum $p$ is given in the units of $\sqrt{\sigma}$; $\sigma$ is the fundamental string tension.}\label{vpplot}
\end{center}
\end{figure}

The selfconsistency condition for the parametrisation (\ref{SiAB}) of the nonlinear Eq.~(\ref{Sigma03}) requires that
the chiral angle is subject to a nonlinear equation --- the mass--gap equation, 
\be 
A_p\cos\vp_p-B_p\sin\vp_p=0, 
\label{mge2} 
\ee 
with
\be
A_p=m+\frac{1}{2}\int\frac{d^3k}{(2\pi)^3}V
(\vec{p}-\vec{k})\sin\vp_k,\quad
B_p=p+\frac{1}{2}\int \frac{d^3k}{(2\pi)^3}\;(\hat{\vec{p}}
\hat{\vec{k}})V(\vec{p}-\vec{k})\cos\vp_k.
\label{AB} 
\ee 
For the given chiral angle $\vp_p$ the dispersive law of the dressed quark can be built as 
\be
E_p=A_p\sin\vp_p+B_p\cos\vp_p, 
\label{Ep} 
\ee 
and it differs drastically from the free--quark energy $E_p^{(0)}=\sqrt{p^2+m^2}$ in the low--momentum domain; 
$E_p$ approaches this free--particle limit as $p\to\infty$. 

It was demonstrated in the pioneering papers on the model (\ref{H}) \cite{Orsay} that, 
for confining potentials, the mass--gap equation (\ref{mge2})
always possesses nontrivial solutions which break chiral symmetry, by generating a nontrivial mass-like function $A_p$, even for
a vanishing quark current mass. 
At Fig.~\ref{vpplot}, as an illustration, we show the numerical solution to the mass--gap equation (\ref{mge2}) 
for the linearly rising potential $V(r)=\sigma r$ in the chiral limit. The chiral angle depicted at Fig.~\ref{vpplot} 
shares all features of chiral symmetry breaking 
solutions to the mass--gap equation (\ref{mge2}) for various confining quark kernels (for a comprehensive analysis of power-like potentials
see Refs.~\cite{Orsay,replica4}), namely, it is
given by a smooth function which starts at $\frac{\pi}{2}$ at the origin,
with the slope inversely proportional to the scale of chiral symmetry breaking, generated by this solution. At large momenta
it approaches zero fast. The latter property allows the reader to anticipate the principal conclusion of this paper. Indeed, since the
Fourier transform of the potential is peaked at $\vpp\simeq\vk$, whereas $\sin\vp_k$ decreases with the increase of $k$,
then, as $p\to\infty$, $A_p$ is a decreasing function of the momentum $p$. In the chiral limit it vanishes asymptotically.
Therefore, for highly excited hadrons, with the typical momentum of valence quarks being large, the dynamical Lorentz--scalar mass
of such valence quarks decreases, and so does the coupling constant $g_\pi$, due to the Goldberger--Treiman relation (\ref{GT}).
Hence highly excited hadrons decouple from the Goldstone bosons. Below we prove this general conclusion by a detailed analysis of the 
amplitude of the pion emission process $h\to h'+\pi$ with $h$ (and perhaps $h'$)
 being highly excited hadrons. 
 
\section{Properties of the Goldstone mode}

\subsection{Mesonic Salpeter vertex; Bethe--Salpeter equation for the chiral pion}

\begin{figure}[t]
\begin{center}
\epsfig{file=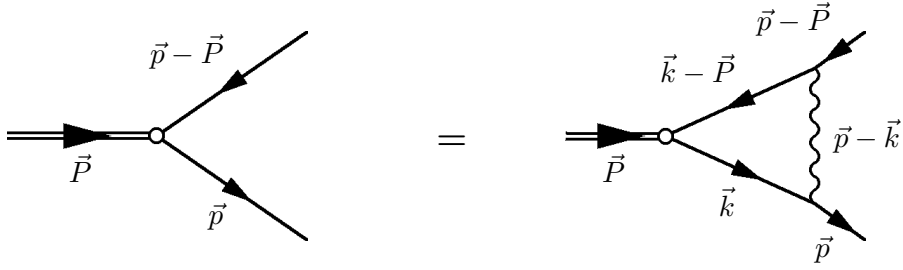,width=12cm} \caption{Graphical representation of the Bethe--Salpeter Eq.~(\ref{GenericSal}) for the
quark--antiquark bound state, in the ladder approximation.}\label{bseq}
\end{center}
\end{figure}

In this chapter we remind the reader the main steps to take in order to derive the Bethe--Salpeter equation for the generic quark--antiquark
bound state, paying special attention to the case of the chiral pion. We follow the lines of Refs.~\cite{BG,Orsay,Lisbon,2d,NR1}.

We start from the homogeneous Bethe--Salpeter equation,
\be
\chi(\vpp,\vec{P})=-i\int\frac{d^4k}{(2\pi)^4}V(\vpp-\vk)\;\gamma_0 S(k_0,\vk)\chi(\vk,\vec{P})S(k_0-M,\vk-\vec{P})\gamma_0,
\label{GenericSal}
\ee
for the mesonic Salpeter amplitude $\chi({\vec p},\vec{P})$; 
$M$ is the mass of the bound state and $\vec P$ is its total momentum. Eq.~(\ref{GenericSal}) is written in the
ladder approximation for the vertex which is consistent with the rainbow approximation for the quark mass operator and which is well
justified in the large-$N_C$ limit. For future convenience we introduce the matrix wave function of the meson as
\be
\Psi(\vec{p},\vec{P})=\int\frac{dp_0}{2\pi}S(p_0,\vpp)\chi(\vec{p},\vec{P})S(p_0-M,\vpp-\vec{P}),
\label{Psi1}
\ee
and use the standard representation for the dressed quark propagator via Dirac projectors,
\be
S(p_0,\vpp)=\frac{\Lambda^{+}({\vec p})\gamma_0}{p_0-E_p+i\epsilon}+ \frac{{\Lambda^{-}}({\vec p})\gamma_0}{p_0+E_p-i\epsilon},
\label{Feynman}
\ee
\be
\Lambda^\pm(\vec{p})=T_pP_\pm T_p^\dagger,\quad
P_\pm=\frac{1\pm\gamma_0}{2},\quad
T_p=\exp{\left[-\frac12(\vec{\gamma}\hat{\vec{p}})\left(\frac{\pi}{2}-\vp_p\right)\right]}.
\label{Tp0}
\ee
We also consider the bound--state in its rest frame setting $\vec{P}=0$ and skipping this argument of the bound--state wave function for
simplicity.
It is easy to perform the integration in energy in Eq.~(\ref{GenericSal}) explicitly. Then, for the Foldy--rotated wave function 
$\tilde{\Psi}(\vec{p})=T^\dagger_p\Psi(\vec{p})T^\dagger_p$, the Bethe--Salpeter Eq.~(\ref{GenericSal}) reads:
\be
\tilde{\Psi}(\vec{p})=-\int\frac{d^3k}{(2\pi)^3}V(\vec{p}-\vec{k})\left[P_+\frac{T_p^\dagger
T_k\tilde{\Psi}(\vec{k})T_kT_p^\dagger}{2E_p-M}P_-+P_-\frac{T_p^\dagger
T_k\tilde{\Psi}(\vec{k})T_kT_p^\dagger}{2E_p+M}P_+\right].
\label{PHI}
\ee
It is clear now that a solution to Eq.~(\ref{PHI}) is to have the form,
\be
\tilde{\Psi}(\vec{p})=P_+{\cal A}P_-+P_-{\cal B}P_+,
\label{AB3}
\ee
and, due to the obvious orthogonality property of the projectors $P_\pm$, $P_+P_-=P_-P_+=0$, only matrices anticommuting with the
matrix $\gamma_0$ contribute to ${\cal A}$ and ${\cal B}$. The set of such matrices is
$\{\gamma_5,\gamma_0\gamma_5,\vec{\gamma},\gamma_0\vec{\gamma}\}$ which can be
reduced even more, down to $\{\gamma_5,\vec{\gamma}\}$,
since the matrix $\gamma_0$ can be always absorbed by the projectors $P_\pm$. The matrix wave
function $\Psi_n(\vpp)$ for the $n$-th generic mesonic state can be parametrised through the positive-- and negative--energy components of the mesonic
wave function $\vp_n^\pm(p)$, and a bound--state equation in the form
\be
\left\{
\begin{array}{l}
[2E_p-M_n]\vp_n^+(p)=\ds\int\frac{\ds k^2dk}{\ds (2\pi)^3}[T^{++}_n(p,k)\vp_n^+(k)+T^{+-}_n(p,k)\vp_n^-(k)]\\[0cm]
[2E_p+M_n]\vp_n^-(p)=\ds\int\frac{\ds k^2dk}{\ds (2\pi)^3}[T^{-+}_n(p,k)\vp_n^+(k)+T^{--}_n(p,k)\vp_n^-(k)]
\end{array}
\right.
\label{bsmes}
\ee
can be derived for such wave functions. The interested reader can find the details in Refs.~\cite{Orsay,Lisbon} or in Ref.~\cite{NR1}, 
where also a Hamiltonian approach to
the quark--antiquark bound--state problem is developed (as a generalisation of the method applied to the 't Hooft model for QCD in two
dimensions in Ref.~\cite{2d}) which, after a generalised Bogoliubov-like transformation leads to the same
Eq.~(\ref{bsmes}); $\vp_n^\pm(p)$ play the role of the bosonic Bogoliubov amplitudes, so that their normalisation condition,
\be
\begin{array}{l}
\ds\int\frac{p^2dp}{(2\pi)^3}\left[\vp_n^{+}(p)\vp_m^{+}(p)-\vp_n^{-}(p)\vp_m^{-}(p)\right]=\delta_{nm},\\[3mm]
\ds\int\frac{p^2dp}{(2\pi)^3}\left[\vp_n^{+}(p)\vp_m^{-}(p)-\vp_n^{-}(p)\vp_m^{+}(p)\right]=0,
\end{array}
\label{normgen}
\ee
should not come as a surprise.

Let us consider the case of the chiral pion in more detail. In this case only $\gamma_5$ contributes to ${\cal A}$ and ${\cal B}$ in
Eq.~(\ref{AB3}) and one has:
\be
{\cal A}_\pi=\gamma_5\vp^+_\pi(p),\quad {\cal B}_\pi=-\gamma_5\vp_\pi^-(p).
\label{AB2}
\ee
It is an easy task now to extract the amplitudes $T_\pi^{\pm\pm}$ (see Eq.~(\ref{bsmes}))
from Eq.~(\ref{PHI}) using the explicit form of the matrix wave function $\tilde{\Psi}_\pi(\vec{p})$ and of the operator $T_p$. 
The result reads:
\be
\begin{array}{lcr}
T_\pi^{++}(p,k)=T_\pi^{-+}(p,k)&=&-\ds\int d\Omega_k V(\vec{p}-\vec{k})
\left[\cos^2\frac{\vp_p-\vp_k}{2}-\frac{1-(\hat{\vec{p}}\hat{\vec{k}})}{2}\cos\vp_p\cos\vp_k\right],\\[5mm]
T_\pi^{+-}(p,k)=T_\pi^{--}(p,k)&=&\ds\int d\Omega_k V(\vec{p}-\vec{k})
\left[\sin^2\frac{\vp_p-\vp_k}{2}+\frac{1-(\hat{\vec{p}}\hat{\vec{k}})}{2}\cos\vp_p\cos\vp_k\right].
\end{array}
\label{pia}
\ee

In the chiral limit, $\vp_\pi^+(p)=-\vp_\pi^-(p)\equiv\vp_\pi(p)$, so that the
bound--state Eq.~(\ref{bsmes}) for the pion reduces to a single equation,
\be
2E_p\vp_\pi(p)=\int\frac{k^2dk}{(2\pi)^3}[T_\pi^{++}(p,k)-T_\pi^{+-}(p,k)]\vp_\pi(k)=
-\int\frac{d^3k}{(2\pi)^3}V(\vec{p}-\vec{k})\vp_\pi(k),
\ee
or, in the coordinate space, one arrives at the Schr{\" o}dinger-like equation,
\be
[2E_p+V(r)]\vp_\pi=0.
\label{bsp1}
\ee
It is instructive to notice that Eq.~(\ref{bsp1}) reproduces the mass--gap Eq.~(\ref{mge2}) for $\vp_\pi(p)=\sin\vp_p$. 
This is the wave function of the chiral pion whose dual nature is clearly seen from this consideration. Indeed,
as a Goldstone boson, the pion appears, through the mass--gap equation, already at the level of the quark dressing, whereas the same 
entity reappears as the lowest pseudoscalar solution to the quark--antiquark bound--state Eq.~(\ref{bsmes}). Below we study the
properties of the pionic matrix wave function (\ref{Psi1}).

\subsection{Bound--state equation in the matrix form: Gell-Mann--Oakes--Renner relation}

Mesonic bound--state Eq.~(\ref{bsmes}) admits a matrix form for the wave function $\Psi(\vpp)$ and it can be derived directly 
from Eq.~(\ref{PHI}). Skipping the details of
this derivation (the interested reader can find them in Ref.~\cite{2d}, for the two--dimensional 't~Hooft model, whereas the 
generalisation to the model (\ref{H}) is trivial), we give it here in the final form \cite{NR1}:
$$
M\Psi(\vec{p})=[(\vec{\alpha}\vec{p})+\beta m]\Psi(\vec{p})+\Psi(\vec{p})[(\vec{\alpha}\vec{p})-\beta m]\hspace*{5cm}
$$
\be
+\int\frac{d^3q}{(2\pi)^3}V(\vec{p}-\vec{k})\left\{\Lambda^+(\vec{k})\Psi(\vec{p})\Lambda^-(-\vec{k})-
\Lambda^+(\vec{p})\Psi(\vec{k})\Lambda^-(-\vec{p})\right.
\label{matrix}
\ee
$$
\hspace*{5cm}\left.-\Lambda^-(\vec{k})\Psi(\vec{p})\Lambda^+(-\vec{k})+\Lambda^-(\vec{p})\Psi(\vec{k})\Lambda^+(-\vec{p})\right\}.
$$
For the chiral pion, the explicit form of $\Psi_\pi(\vec{p})$ follows from Eqs.~(\ref{AB3}) and (\ref{AB2}) and reads:
\be
\Psi_\pi(\vec{p})=T_p\left[P_+\gamma_5\vp_\pi^+-P_-\gamma_5\vp_\pi^-\right]T_p=\gamma_5G_\pi+\gamma_0\gamma_5T_p^2F_\pi,
\label{exp}
\ee
where $G_\pi=\frac12(\vp_\pi^+-\vp_\pi^-)$, $F_\pi=\frac12(\vp_\pi^++\vp_\pi^-)$. 

In order to normalise the pion wave function in its rest frame, one is to go slightly beyond the chiral limit and to 
consider pionic solutions to the bound--state Eq.~(\ref{bsmes}) in the form \cite{Lisbon,2d,NR1}:
\be
\vp_\pi^\pm(p)={\cal N}_\pi\left[\pm\frac{1}{\sqrt{m_\pi}}\sin\vp_p+
\sqrt{m_\pi}\Delta_p\right],\quad
{\cal N}_\pi^{-2}=4\int_0^\infty\frac{p^2dp}{(2\pi)^3}\Delta_p\sin\vp_p,
\label{vppm}
\ee
where all corrections of higher order in the pion mass are neglected and the function
$\Delta_p$ obeys a reduced $m_\pi$--independent equation
(see, for example, Ref.~\cite{Lisbon} or the papers \cite{2d} where such an
equation for $\Delta_p$ is discussed in two-dimensional QCD).

Furthermore, the pion norm ${\cal N}_\pi$ can be easily related to the pion decay constant $f_\pi$. To this end we 
multiply the matrix bound--state equation (\ref{matrix}) by $\gamma_0\gamma_5$, integrate its
both parts over $\frac{d^3p}{(2\pi)^3}$, and, finally, take the trace. The resulting equation reads:
\be
m_\pi\int\frac{d^3p}{(2\pi)^3}F_\pi\sin\vp_p=2m\int\frac{d^3p}{(2\pi)^3}G_\pi,
\label{GMOR2}
\ee
and it is easy to recognise the celebrated Gell-Mann--Oakes--Renner relation \cite{GMOR} in Eq.~(\ref{GMOR2}). Indeed,
using the explicit form of the pionic wave function beyond the chiral limit, Eq.~(\ref{vppm}), one can
see that $G_\pi=({\cal N}_\pi/\sqrt{m_\pi})\sin\vp_p$ and $F_\pi=({\cal N}_\pi\sqrt{m_\pi})\Delta_p$ which, after substitution to
Eq.~(\ref{GMOR2}), give the sought relation,
\be
m_\pi^2\left[\frac{N_C}{\pi^2}\int_0^\infty dp\; p^2\Delta_p\sin\vp_p\right]=-2m\langle\bar{q}q\rangle,
\label{GMOR3}
\ee
where the definition of the chiral condensate \cite{Orsay,Lisbon},
\be
\langle\bar{q}q\rangle=-\frac{N_C}{\pi^2}\int^{\infty}_0 dp\;p^2\sin\vp_p,
\label{Sigma1}
\ee
was used. Therefore,
\be
{\cal N}_\pi=\frac{\sqrt{2\pi N_C}}{f_\pi}.
\ee

Finally, 
\be
\Psi_\pi(\vpp)=T_p\left[P_+\gamma_5\vp_\pi^+(p)-P_-\gamma_5\vp_\pi^-(p)\right]T_p=
\frac{1}{f_\pi}\sqrt{\frac{2\pi N_C}{m_\pi}}\bigl[\gamma_5\sin\vp_p+O(m_\pi)\bigr],
\label{psip}
\ee
where the properties $T_p\gamma_5=\gamma_5T_p^\dagger$ and $T_p^\dagger T_p=T_pT_p^\dagger=1$ of the Foldy operator $T_p$ 
(see Eq.~(\ref{Tp0}) for its definition) were taken into account. 

\subsection{Goldberger--Treiman relation and the pion emission vertex}

The standard Goldberger--Treiman relation is the relation between the pion--nucleon coupling constant and the axial constant of the nucleon
axial vector current, and its derivation is present in any textbook on hadronic physics (see, for example, Ref.~\cite{CL}). This relation
can be derived analytically for the GNJL model as well \cite{BicudAp}.
A similar relation holds in the GNJL models for the pion coupling to dressed quarks. Indeed, the
model admits two representations: the dressed--quark and the mesonic representation which are interchangeable up to 
corrections suppressed by the large $N_C$ number \cite{NR1}, so that, in the leading in the number of colours approximation, one has two
equivalent representations for the axial--vector current complying with the PCAC theorem (for the sake of simplicity, we
stick to the one--flavour theory and ignore the axial anomaly, generalisation to
the flavour nonsinglet axial vector current in $SU(N_f)$, where there is no anomaly, 
is trivial)\ftnote{1}{It was noticed long ago
\cite{Orsay} that, in the given model, the pion decay constant in the temporal and in the spatial parts of the current 
in Eq.~(\ref{piax}) may differ. This is a consequence of an explicit breaking of Lorentz covariance by the instantaneous interaction in
the Hamiltonian (\ref{H}). Although some improvements of the model can be made in order to get rid of this discrepancy 
(see, for example, Ref.~\cite{Bicudo}), its underlying origin cannot be removed by simple amends. Thus we consider the model (\ref{H}) as it is, making
emphasis on its qualitative predictions. Everywhere throughout this paper as $f_\pi$ we denote the temporal constant extracted from the
Gell-Mann--Oakes--Renner relation (\ref{GMOR3}).}:
\be
[J_\mu^5(x)]_\pi=f_\pi\partial_\mu \phi_\pi(x),
\label{piax}
\ee
through the Goldstone boson, with $\phi_\pi(x)$ being the chiral pion wave function, and, generically,
\be
[J_\mu^5(x)]_q=\bar q(x)[g_A \gamma_\mu \gamma_5+h_A\partial_\mu \gamma_5]q(x),
\label{qax}
\ee
through the dressed--quark states, where $g_A=1$ and $h_A=0$. We evaluate now the matrix element of this current divergence between dressed 
quark states, $\langle q(p)|\partial_\mu J_\mu^5(x)|q(p')\rangle$, in the given two representations arriving at:
\be
\langle q(p)|[\partial_\mu J_\mu^5(x)]_\pi|q(p')\rangle=f_\pi m_\pi^2 \langle q(p)|\phi_\pi(x)|q(p')\rangle
\propto f_\pi g_\pi(q^2) (\bar{u}_p\gamma_5 u_{p'}),\quad q=p-p',
\label{eq1}
\ee
where the pion--quark--quark effective formfactor $g_\pi(q^2)$ is introduced, and
\be
\langle q(p)|[\partial_\mu J_\mu^5(x)]_q|q(p')\rangle\propto m_q^{\rm eff}(\bar{u}_p\gamma_5 u_{p'}),
\label{eq2}
\ee
where it is assumed that dressed quarks obey an effective Dirac equation with the dynamically generated 
effective  Lorentz--scalar mass $m_q^{\rm eff}$. Obviously this dynamical
mass must be identified with the function $A_p$ --- see Eqs.~(\ref{SAB}) and (\ref{AB}), --- and we discuss this issue in detail below.

Finally, Eqs.~(\ref{eq1}) and (\ref{eq2}) together yield 
the needed relation,
\be
f_\pi g_\pi=m_q^{\rm eff},\quad g_\pi\equiv g_\pi(m_\pi^2),
\label{GTr}
\ee
where, for the sake of simplicity, we absorb all coefficients into the definition of $g_\pi$. 

This  derivation of the
Goldberger--Treiman relation (\ref{GTr}) is based on quite general
considerations of chiral symmetry breaking and the PCAC theorem.
The only input is the requirement of the (partial) conservation
of the total axial vector current in QCD and appearance of
the Lorentz--scalar dynamical mass of quarks as a consequence
of SBCS. It leaves a number of questions,
such as the microscopic picture for the pion--quark--quark vertex and 
the dependence of the effective quark mass
$m_q^{\rm eff}$ on the quark momentum. Below we give a microscopic derivation of the Goldberger--Treiman-like relation (\ref{GTr}) for the
pion--quark--quark vertex in the framework of the quark model (\ref{H}).

\begin{figure}[t]
\begin{center}
\epsfig{file=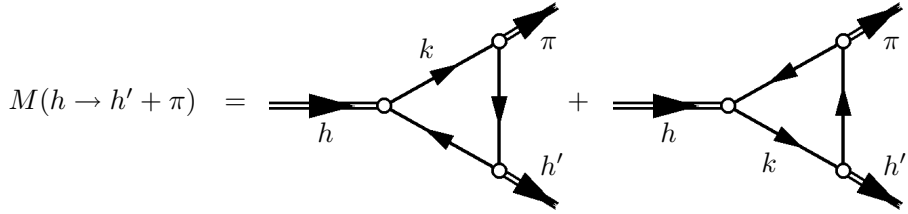,width=12cm} 
\caption{The amplitude of the decay $h\to h'+\pi$.}\label{hhpi}
\end{center}
\end{figure}

Consider a hadronic process with an emission of a soft pion, for example, a decay $h\to h'+\pi$, with $h$ and $h'$ being hadronic
states (below we consider them to be mesons) with the total momenta $\vpp$ and $\vpp'$, respectively. 
The amplitude of this process is given by the sum of two triangle diagrams depicted at Fig.~\ref{hhpi} and can be written as
\be
\begin{array}{c}
\ds M(h\to h'+\pi)=\int\frac{d^4k}{(2\pi)^4}Sp\left[\chi_h(\vk,\vpp)S(k-p)\bar{\chi}_{h'}(\vk-\vpp,\vpp')
S(k-q)\bar{\chi}_\pi(\vk,\vq)S(k)\right]\\[3mm]
\ds +\int\frac{d^4k}{(2\pi)^4}Sp\left[\chi_h(\vk,\vpp)S(k-p)\bar{\chi}_\pi(\vk,\vq)S(k-q)\bar{\chi}_{h'}(\vk-\vpp,\vpp')
S(k)\right],
\end{array}
\label{Mhhpi}
\ee 
where the pion momentum is $q=p-p'$.
Every vertex in these diagrams contains a Salpeter amplitude $\chi$ ($\bar{\chi}$ for outgoing vertices)
which obeys a Bethe--Salpeter equation of the form of Eq.~(\ref{GenericSal}) which can be rewritten as
\be
\chi(\vpp,\vec{P})=\int\frac{d^3k}{(2\pi)^3}V(\vpp-\vk)\gamma_0\Psi(\vk,\vec{P})\gamma_0.
\label{bsp2}
\ee
The incoming and outgoing matrix vertices for the given hadron are related to one another as 
\be
\bar{\chi}(\vpp,\vec{P})=\gamma_0\chi^\dagger(\vpp,\vec{P})\gamma_0.
\label{chib}
\ee

Thence, considering Eqs.~(\ref{psip}), (\ref{bsp2}), and (\ref{chib}) together, one arrives at the relation between the soft pion 
($\vec{P}_\pi\equiv \vq\to 0$) emission vertex and the dressed quark effective dynamical mass:
\be
f_\pi\bar{\chi}_\pi(\vpp)=\sqrt{\frac{2\pi N_C}{m_\pi}}\gamma_5\int\frac{d^3k}{(2\pi)^3}V(\vpp-\vk)\sin\vp_k=
{\rm const}\times\gamma_5 A_p,
\label{Ap}
\ee
where the definition of the function $A_p$, Eq.~(\ref{AB}), was used. Thus, for the sake of transparency, introducing the formfactor 
$g_\pi(p)$ such that
\be
\bar{u}_p\bar{\chi}_\pi(\vpp)u_{p}={\rm const}\times g_\pi(p)(\bar{u}_p\gamma_5u_{p}),
\ee
with the same constant as in Eq.~(\ref{Ap}), one finally arrives at the Goldberger--Treiman relation,
\be
f_\pi g_\pi(p)=A_p,
\label{GTr2}
\ee
which explicitly relates the pion coupling to the dressed quarks with the effective chiral symmetry breaking Lorentz--scalar
quark mass $A_p$. Notice an important difference between the naive Goldberger--Treiman 
relation (\ref{GTr}) and the microscopic relation (\ref{GTr2}). In the former case, the pion coupling constant depends only of the momentum transfer in the pionic
vertex (that is, the pion total momentum $\vq$) squared, so that, for the on--shall pion, $g_\pi$ is a constant. 
On the contrary, the pion emission vertex $\bar{\chi}_\pi(\vpp,\vq)$ in the microscopic GNJL model
depends on two arguments: one, as before, being the pion total momentum
$\vq$ --- as mentioned before, we treat it as for the standard derivation of the
Goldberger--Treiman relation and continue the soft--pion vertex to the
point $\vq=0$ --- whereas the other argument being the momentum flow
in the loop, which plays also the role of the momentum of the dressed quark the pion couples to.
Without SBCS and beyond the chiral limit, $A_p=m$ and thus  the coupling $g_\pi(p)$ turns to a constant.
On the contrary, in the chiral limit and after SBCS, $A_p$ is a rapidly decreasing function of the momentum $p$. Indeed, using the definition
of $A_p$ and that of the dressed quark dispersive law, one can find the relation:
\be
A_p=E_p\sin\vp_p. 
\label{App}
\ee
At large momenta, $E_p\approx p$, whereas the chiral angle approaches zero fast --- see Fig.~\ref{vpplot}. 
For a power-like confining potential (\ref{potential}) $\vp_p\propto 1/p^{4+\alpha}$ as $p\to\infty$ (see, for example, 
Ref.~\cite{replica4}).

The decay amplitude of the process $h\to h'+\pi$ is given by an overlap of the three vertices --- see Fig.~\ref{hhpi} and Eq.~(\ref{Mhhpi}). 
Each of them depends on the momentum circulating in the loop, and the maximal overlap is achieved for all three meson wave functions 
localised at comparable values of this momentum. Clearly the pion vertex (and the pion wave function $\sin\vp_p$) is dominated
by the low momenta and it decreases fast with the increase of the latter (see Fig.~\ref{vpplot}). In the meantime, 
for highly excited hadrons the momentum distribution in these hadrons as well as in the corresponding vertices
in Fig.~\ref{hhpi} is shifted to large momenta. Therefore, the amplitude (\ref{Mhhpi}) vanishes
with the increase of the hadron $h$ or/and $h'$ excitation number and so does the effective coupling
constant of the pion to such highly excited hadrons. We emphasise that it is the pion wave function $\sin\vp_p$ that suppresses the
Goldstone boson coupling to highly excited hadrons.

It was demonstrated recently that chiral symmetry restoration for excited hadrons happens due to the same reason --- the effective
interaction responsible for the splitting within a chiral doublet is also proportional to $\sin\vp_p$ \cite{parity2}. 
Therefore, if an effective constant is introduced in particular for the pion transition within the approximate parity doublet, 
this constant must be proportional to the splitting $\Delta M_{+-}=M_+ -M_-$ within the doublet and it has to vanish with 
$\Delta M_{+-}$ for highly excited hadrons.

\subsection{Numerical estimates}

\begin{figure}[t]
\begin{center}
\epsfig{file=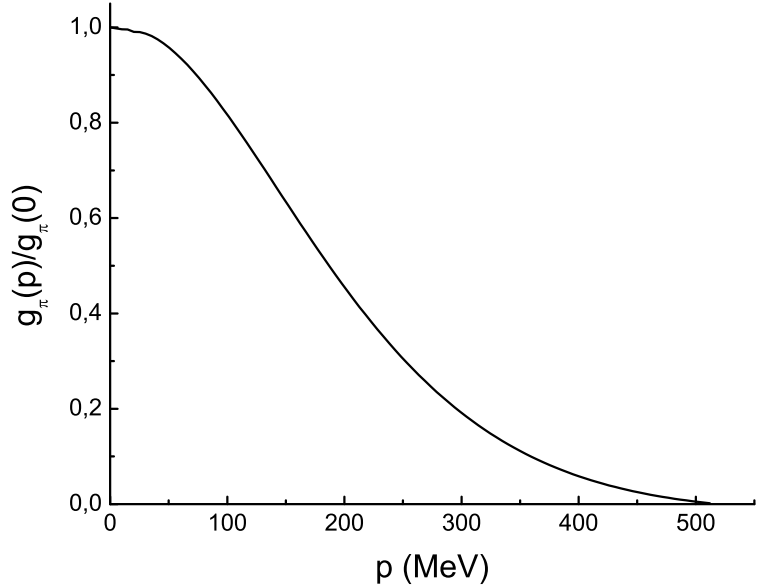,width=10cm} 
\caption{The ratio $g_\pi(p)/g_\pi(0)$ as a function of the dressed quark momentum.}\label{gpiplot}
\end{center}
\end{figure}

In this chapter we present some quantitative estimates related to the chiral pion and its coupling to excited hadrons.
For the sake of simplicity, we stick to the harmonic oscillator potential corresponding to the marginal choice of $\alpha=2$ in Eq.~(\ref{potential}).
The mass--gap equation reduces in this case to a second--order differential equation studied in detail in Refs.~\cite{Orsay,Orsay2,Lisbon},
\be
p^3\sin\vp_p=\frac12K_0^3\left[p^2\vp''_p+2p\vp_p'+\sin2\vp_p\right]+mp^2\cos\vp_p,
\label{diffmge}
\ee
where the dressed quark dispersive law is
\be
E_p=m\sin\vp_p+p\cos\vp_p-K_0^3\left[\frac{{\vp'_p}^2}{2}+\frac{\cos^2\vp_p}{p^2}\right].
\label{Epharm}
\ee

The bound--state equation for the chiral pion, Eq.~(\ref{bsmes}), takes the form
\be
\left[ \left[-K_0^3\frac{d^2}{dp^2}+2E_p\right]\left[
\begin{array}{cc}
1&0\\0&1
\end{array}
\right]
+K_0^3
\left[
\frac{\vp^{'2}_p}{2}+\frac{\cos^2\vp_p}{p^2}\right]\left[
\begin{array}{cc}1&1\\1&1\end{array}
\right]
-
m_\pi\left[
\begin{array}{cc}1&0\\0&-1\end{array}
\right]
\right]
\left[
\begin{array}{c}
\nu_\pi^+(p)\\
\nu_\pi^-(p)
\end{array}
\right]=0, 
\label{hop}
\ee
where the radial wave functions $\nu^\pm_\pi(p)=p\vp^\pm_\pi(p)$ were introduced for convenience,
which are rescaled so as to obey the one-dimensional normalisation \cite{Lisbon},
\be
\int dp[\nu^{+2}_\pi(p)-\nu^{-2}_\pi(p)]=1.
\ee  
As noticed before, Eq.~(\ref{hop}) reduces to the mass--gap Eq.~(\ref{diffmge}) in the chiral limit, so that $\vp_\pi^\pm(p)=\sin\vp_p$
and the qualitative form of the solution for the chiral angle appears to be the same as for the linear confinement depicted in Fig.~\ref{vpplot}.

In Fig.~\ref{gpiplot}, we plot the ratio $g_\pi(p)/g_\pi(0)$ defined with the
 help of Eqs.~(\ref{GTr2}) and (\ref{App}). This ratio also describes the 
actual decrease of the 
pion coupling to hadron with the increase of the average dressed quark 
momentum in it. Furthermore, for the sake of transparency, we measure this momentum in 
physical units, in $MeV$, using the chiral condensate to fix the value of the mass parameter $K_0$. To this end, for the given solution to the mass--gap
Eq.~(\ref{diffmge}), we calculate the chiral condensate according to Eq.~(\ref{Sigma1}) which gives (we use the numerical results of Refs.~\cite{NR1,parity2})
\be
\langle\bar{q}q\rangle=-(0.51K_0)^3
\ee
and, therefore, we fix $K_0=490MeV$ to arrive at the standard value of the chiral condensate. From Fig.~\ref{gpiplot} one can see a fast decrease of the pion coupling to hadrons
for large $p$'s, as was discussed before using general qualitative arguments. 

Now we estimate the average momentum of the valence quarks in excited mesons.
We consider a radially excited $S$-wave light--light meson consisting of two light (massless) quarks.
For highly excited bound states of such a system ($n\gg 1$), the negative--energy component of the wave function $\vp_n^-(p)$ vanishes and the bound--state equation reduces to a
single equation for the positive--energy component $\vp_n^+(p)$. This equation can be roughly approximated by the Schr{\" o}dinger equation with the Hamiltonian
\be
H=2|p|+K_0^3r^2,
\label{rp1}
\ee
and we identify the corresponding selfenergies $M_n$ as $M_n=\frac12(M_{n+}+M_{n-})$, with $M_{n\pm}$ being the masses for the chiral doublet partners. 
Although the given naive Salpeter Hamiltonian is unable to correctly discriminate between $M_{n+}$ and $M_{n-}$ and a more sophisticated approach is required for this purpose 
(see Ref.~\cite{parity2} for a detailed discussion), it is sufficient to 
estimate the typical quark momentum in excited states. To make things simpler, we use the einbein field method
(see the original paper \cite{ein1} for the details and, for example, Ref.~\cite{ein2} for einbeins treated using the Dirac constraints formalism \cite{Dirac}) rewriting
the Hamiltonian (\ref{rp1}) as
\be
H=\frac{p^2}{\mu}+\frac{\mu}{2}+K_0^3r^2,
\label{rp2}
\ee
with the einbein $\mu$ treated as a variational parameter \cite{ein3}. The spectrum of the Hamiltonian (\ref{rp2}) minimised with respect to the parameter $\mu$ reads
$M_n\approx 3K_0(2n+3/2)^{2/3}$ and the mean quark momentum can be easily estimated then using the virial theorem for the oscillator Hamiltonian (\ref{rp2}) to be
\be
\langle p\rangle\approx K_0 \left(2n+\frac32\right)^{2/3},
\ee
which gives an approximate analytic dependence of $\langle p\rangle$ on the radial excitation number $n$. 
It is obvious therefore that already for $n\sim 1$ the quark momentum appears large enough and, as seen from Fig.~\ref{gpiplot}, 
the corresponding coupling of the Goldstone boson to this hadron is suppressed as compared to its naive value of $g_\pi(0)$. An accurate systematic evaluation of the
pion coupling to excited hadrons goes beyond the scope of the present qualitative paper --- this work is in progress now and will be reported elsewhere.

\section{Conclusions}

In this paper we considered in detail physics of the Goldstone bosons decoupling
from highly excited hadrons, where chiral symmetry is approximately restored. This interesting physics
can be summarised in a few words. The coupling of the Goldstone bosons
to the valence quarks is regulated, via the Goldberger--Treiman relation, by the dynamical Lorentz--scalar mass of quarks.
This dynamical mass which arises via the loop dressing of quarks represents
effects of chiral symmetry breaking in the vacuum. A key feature
of this dynamical mass is that it is strongly momentum--dependent and
vanishes at large quark momenta. Hence at large momenta the valence
quarks decouple from the quark condensates and from the
Goldstone bosons. In this regime chiral symmetry breaking
in the vacuum almost does not affect observables and physics is essentially such
as if there had been no such chiral symmetry breaking in the vacuum.
It is clear, therefore, that for a given hadron, the influence of SBCS on its properties is determined by the typical
momentum of valence quarks in it. For low--lying hadrons this momentum lies below the chiral symmetry breaking scale,
the valence quarks acquire a significant Lorentz--scalar dynamical mass,
which results in a strong coupling of the low--lying hadrons to the Goldstone bosons. On the contrary, for high--lying hadrons,
a typical momentum of valence quarks is large, their dynamical chiral symmetry breaking mass becomes small (and asymptotically
vanishes), which implies that these high--lying hadrons decouple from the Goldstone bosons. Consequently the
axial vector constant, $g_A$, of the highly excited hadrons appears suppressed and asymptotically vanishes.

We illustrate this physics in the framework of the GNJL model for QCD
with the only interaction between quarks being the instantaneous Lorentz--vector confining potential. Such a model is tractable
and contains all the required elements such as the spontaneous breaking of chiral symmetry via quantum fluctuations of the
quark fields and confinement. While being only a model, it nevertheless
provides a significant insight into physics of chiral symmetry restoration in excited hadrons.

\begin{acknowledgments}
L. Ya. G. thanks A. Gal for his hospitality
at the Hebrew University of Jerusalem where this paper
was prepared, and T. Cohen, M. Shifman and A. Vainstein for
correspondence. He acknowledges the support of the P16823-N08 project
of the Austrian Science Fund. A. V. N. would like to 
thank Yu. S. Kalashnikova and E. Ribeiro for reading the manuscript and
critical comments and to acknowledge 
support of the grants DFG 436 RUS 113/820/0-1, RFFI 05-02-04012-NNIOa, NSh-843.2006.2
as well as of the Federal Programme of the Russian Ministry of Industry, 
Science, and
Technology No 40.052.1.1.1112.
\end{acknowledgments}

\end{document}